\documentstyle[12pt]{article}

\newcommand {\vs}[1]  { \vspace*{#1 cm} }

\newcounter{eq}
\newcounter{sc}

\newcommand {\AP}   {Ann. of Phys.}
\newcommand {\CQG}  {Class. Quantum. Grav.}

\newcommand {\IJMP}  {Int. J. Mod. Phys.}
\newcommand {\JHEP}   {JHEP}
\newcommand {\MPL}  {Mod. Phys. Lett.}
\newcommand {\NP}   {Nucl. Phys.}

\newcommand {\PL}   {Phys. Lett.}
\newcommand {\PR}   {Phys. Rev.}
\newcommand {\PRL}   {Phys. Rev. Lett.}
\newcommand {\PTP}  {Prog. Theor. Phys.}

\def\overleftrightarrow#1{\vbox{\ialign{##\crcr
 $\leftrightarrow$\crcr\noalign{\kern-1pt\nointerlineskip}
 $\hfil\displaystyle{#1}\hfil$\crcr}}}

\newlength{\minitwocolumn}
\begin{document}

\begin{flushright}
DPUR/TH/17\\
May, 2009\\
\end{flushright}
\vspace{30pt}
\pagestyle{empty}
\baselineskip15pt

\begin{center}
{\large\bf Renormalizability of Topologically Massive Gravity
 \vskip 1mm
}

\vspace{20mm}

Ichiro Oda
          \footnote{
           E-mail address:\ ioda@phys.u-ryukyu.ac.jp
                  }

\vspace{10mm}
          Department of Physics, Faculty of Science, University of the 
           Ryukyus,\\
           Nishihara, Okinawa 903-0213, JAPAN \\

\end{center}


\vspace{20mm}
\begin{abstract}
We consider renormalizability of topologically massive gravity in three space-time dimensions.  
With a usual parametrization of the metric tensor, we establish the statement that topologically 
massive gravity is in fact renormalizable. In this proof, we make use of not only
a recently found, new infrared regularization method of scalar mode but also a covariant ultraviolet 
regulator with a specific combination of higher derivative terms which is motivated by the
new massive gravity in three dimensions. 
\vspace{15mm}

\end{abstract}

\newpage
\pagestyle{plain}
\pagenumbering{arabic}


\rm
\section{Introduction}
In recent years there has been a revival of interest in three-dimensional
quantum gravity mainly owing to the paper by Witten \cite{Witten}
and that by Li, Song and Strominger \cite{Strominger}. Both the papers deal with 
not only the AdS/CFT correspondence in three dimensions but also the problem
of counting the microscopic degrees of freedom associated with BTZ
black holes by utilizing certain holographic dual theories at a boundary
which are two-dimensional. Even if three-dimensional quantum gravity
is considered in these papers, Witten investigates topological AdS gravity 
whereas Li et al. studies, what is called, "topologically massive gravity (TMG)".
In this article, we shall focus on the problem of renormalizability
of the latter three-dimensional gravity.

Topologically massive gravity (TMG) \cite{Deser1, Deser2} in three space-time dimensions is 
described by the action consisted of the Einstein-Hilbert action with
$\it{wrong}$ sign and gravitational Chern-Simons term which is parity-violating
and includes three derivatives, and has one dynamical degree of freedom corresponding 
to massive graviton of $+2$ or $-2$ helicity mode depending on the sign of 
the overall constant in front of gravitational Chern-Simons term.
It is remarkable that despite the presence of three derivatives in gravitational 
Chern-Simons term, there are neither ghosts nor acausalities in TMG. 

Moreover, it is expected that the higher derivative term in TMG would dominate 
the behavior of the theory at high energies, leading to a stabilization of the 
ultraviolet divergence and consequently to power-counting renormalizability.
Actually, the problem of renormalizability of TMG has been discussed before,
but we have not yet had a firm grasp of it since we have no gauge-invariant 
regularization method in such a way to preserve the desirable power-counting behavior. 
This absence of the useful ultraviolet regulator is of course related to the fact
that TMG includes the Levi-Civita tensor density $\varepsilon^{\mu\nu\rho}$ in the
classical action so that we cannot make use of dimensional regularization. Furthermore, 
it turns out that gauge-invariant, higher derivative regulators spoil the argument 
of formal renormalizability, non-covariant cutoffs cannot be easily analyzed, 
and non-local regularization method involves some assumption to be proved \cite{Deser3, Kleppe}. 

Recently, in three space-time dimensions there has been an interesting progress 
for obtaining a sensible interacting massive gravity theory \cite{Bergshoeff1, Bergshoeff2}
\footnote{See the references \cite{Percacci, Kaku1, Porrati, Kirsch, 
't Hooft, Kaku2, Oda1, Maeno1, Maeno2} for alternative massive gravity models.}. 
This model has been shown to be equivalent to the Pauli-Fierz massive gravity \cite{Fierz}
at the linearized approximation level and thus massive modes of helicities $\pm 2$ 
are physical propagating ones \footnote{Recall that the massive graviton in the Pauli-Fierz 
theory possesses $\frac{(D+1)(D-2)}{2}$ propagating modes in a general $D$ dimension.}. 
A key idea in this model is that one adds a specific combination of higher derivative curvature terms 
to the Einstein-Hilbert action with the $\it{wrong}$ sign in such a way that the trace part 
of the stress-energy tensor associated with those higher derivative terms is proportional to 
the original higher derivative Lagrangian. 
With this idea, it turns out that the scalar mode coming from higher derivative Lagrangian
is precisely cancelled out \cite{Nakasone2} and consequently we have a conformal invariance
at least in the higher derivative sector (Of course, the Einstein-Hilbert action breaks
the conformal invariance). Afterwards, this new massive gravity model in three 
dimensions has been studied from various viewpoints such as the unitarity and the impossibility 
of generalization to higher dimensions \cite{Nakasone1}, relation to the Pauli-Fierz mass 
term \cite{Nakasone2}, black hole solutions \cite{Clement1, Clement2}, the properties of 
linearized gravitational excitations in asymptotically AdS space-time \cite{Liu1, Liu2, Liu3}, 
AdS waves \cite{Eloy}, the z=4 Horava-Lifshitz Gravity \cite{Cai}, the no-go theorem \cite{Deser4}
and the effects of torsion \cite{Hernaski}. 

More recently, we have presented a proof that the new massive gravity theory in three dimensions
\cite{Bergshoeff1, Bergshoeff2} is certainly renormalizable \cite{Oda2}. In this proof, 
we have used a particular BRST-invariant infrared regularization procedure for the scalar mode \cite{Oda2}. 
This new regularization method is composed of two steps. The first step is to add the conventional 
Pauli-Fierz mass term in the new massive theory and then to make the Pauli-Fierz mass term 
BRST-invariant by applying the Stueckelberg formalism. The next step is to take the massless limit 
after renormalization procedure. Incidentally, the procedure of making the BRST-invariant Pauli-Fierz mass 
term was previously considered by Hamamoto \cite{Hamamoto}, in which the motivation was to construct a massive 
tensor theory with a smooth massless limit.  

It is then natural to ask ourselves if this infrared regularization method could be also applied to 
topologically massive gravity (TMG) in three dimensions or not. The purpose of this article is
to show that the answer to this question is affirmative. However, in the process we will soon realize 
that compared with the new massive gravity \cite{Bergshoeff1, Bergshoeff2}, we encounter a new difficulty which 
amounts to the impossibility of making use of dimensional regularization in TMG since there is 
the Levi-Civita tensor density $\varepsilon^{\mu\nu\rho}$ in gravitational Chern-Simons term. 
Note that this difficulty is related to the problem what regularization method for the ultraviolet 
divergence we should adopt. In this article, we shall adopt a diffeomorphism-invariant, higher derivative 
regularization method. A peculiar feature of this regulator is that the combination of the higher 
derivative terms shares the same structure as that of the new massive gravity, by which we have no
scalar mode in graviton propagator obtained by the regulator terms.

In the next section, we briefly review on topologically massive gravity. In 
the third section, we explain how to construct the BRST-invariant Pauli-Fierz massive
gravity via the Stueckelberg formalism. 
In the fourth section, we derive the propagator of the gravitational field on the
basis of the gauge fixed, BRST-invariant action obtained in the section 3.
In the fifth section, we calculate the superficial degree of divergence and present the
ultraviolet regulator.
The final section is devoted to conclusion and discussions.

\section{Brief review of topologically massive gravity}

We start with brief review of topologically massive gravity in three space-time
dimensions \cite{Deser1, Deser2}. The action takes the form
\begin{eqnarray}
S_c &\equiv& \int d^3 x {\cal{L}}_c
\nonumber\\
&=& \int d^3 x [ - \frac{1}{\kappa^2} \sqrt{- g} R  
+ \frac{1}{2 \kappa^2 \mu} \varepsilon^{\lambda\mu\nu} \Gamma_{\lambda\sigma}^\rho
( \partial_\mu \Gamma_{\rho\nu}^\sigma + \frac{2}{3} \Gamma_{\mu\tau}^\sigma
\Gamma_{\nu\rho}^\tau ) ],
\label{TMG}
\end{eqnarray}
where $\kappa^2 \equiv 16 \pi G$ ($G$ is the $3$-dimensional Newton's constant) and
$\mu$ is a constant of mass dimension.
Let us note that $\kappa$ has dimension of $(mass)^{-\frac{1}{2}}$, so
the theory defined by the action (\ref{TMG}) might at first sight appear to be
unrenormalizable, but it turns out to be an illusion. The space-time indices 
$\mu, \nu, \cdots$ run over $0, 1, 2$, we take the metric signature $(-, +, +)$, 
and follow the notation and conventions of the textbook of MTW \cite{MTW}. Finally, 
the Levi-Civita tensor density is defined as $\varepsilon^{012} = 1$. 

Variation of the classical action (\ref{TMG}) yields the equations of motion
\begin{eqnarray}
G^{\mu\nu} - \frac{1}{\mu} C^{\mu\nu} = 0,
\label{Eq. of motion}
\end{eqnarray}
where $C^{\mu\nu}$ is the Cotton conformal tensor density defined by
\begin{eqnarray}
C^{\mu\nu} &=& \frac{1}{2 \sqrt{-g}} (\varepsilon^{\mu\alpha\beta} 
\nabla_\alpha R_\beta ^\nu + \varepsilon^{\nu\alpha\beta} 
\nabla_\alpha R_\beta ^\mu)
\nonumber\\
&=& \frac{1}{\sqrt{-g}} \varepsilon^{\mu\alpha\beta} 
\nabla_\alpha (R_\beta ^\nu - \frac{1}{4} \delta_\beta^\nu R).
\label{Cotton}
\end{eqnarray}
In deriving the second equality, we have used an identity
\begin{eqnarray}
- \varepsilon^{\mu\alpha\beta} \nabla_\alpha R_\beta ^\nu 
+ \varepsilon^{\nu\alpha\beta} \nabla_\alpha R_\beta ^\mu
= \frac{1}{2} \varepsilon^{\mu\nu\alpha} \nabla_\alpha R,
\label{Id}
\end{eqnarray}
which is obtained by using both the Bianchi identity
\begin{eqnarray}
\nabla_{[\mu} R_{\nu\rho]\sigma} \ ^\lambda = 0, 
\label{Bianchi}
\end{eqnarray}
and the relation holding only in three dimensions 
\begin{eqnarray}
R_{\mu\nu} \ ^{\rho\sigma}  = 4 \delta_{[\mu} ^{[\rho} R_{\nu]} ^{\sigma]}
-  \delta_{[\mu} ^{\rho} \delta_{\nu]} ^{\sigma} R, 
\label{3D relation}
\end{eqnarray}
where the square bracket denotes the antisymmetrization of indices with
a numerical weight. Note that the Cotton tensor density has the properties
$C^\mu \ _\mu = 0, C^{\mu\nu} = C^{\nu\mu}$, and $\nabla_\mu C^{\mu\nu} = 0$.

It turns out that the linearized equations of motion are those of a massive
scalar field
\begin{eqnarray}
( \Box + \mu^2 ) \phi = 0, 
\label{KG}
\end{eqnarray}
where 
\begin{eqnarray}
\phi = (\delta_{ij} + \hat{\partial_i} \hat{\partial_j}) h^{ij}, 
\label{phi}
\end{eqnarray}
with the definitions of $\Box = \eta^{\mu\nu} \partial_\mu \partial_\nu, 
\hat{\partial_i} = \frac{\partial_i}{\sqrt{- \nabla^2}}$ and $i, j, \cdots = 1, 2$. 
The existence of such a massive mode is also confirmed by examining the effective
force between two external gravitational sources and finding the Yukawa-type 
interaction. Here it should be emphasized that despite the Klein-Gordon equation (\ref{KG})
the graviton carries spin either $+2$ or $-2$ (parity-violating because of the
presence of the Levi-Civita tensor density) and the spin sign is correlated
with that of the coefficient $\mu$ in front of gravitational Chern-Simons
term.

One important remaining problem in TMG is to prove that this theory is perturbatively
renormalizable. If so, TMG would give us a scarce example of consistent quantum gravity 
theories in lower dimensions and bring about a better underdstanding of how to 
construct a satisfactory quantum gravity in four dimensions in future.
Indeed, this problem has been attacked by two groups \cite{Deser3, Kleppe},
but still unsolved completely.  The reason why the problem is difficult is
quite simple. Although there is an improvement in the ultraviolet behavior owing to 
gravitational Chern-Simons term with three derivatives and TMG consequently seems to be 
manifestly power-counting renormalizable, there is a nasty problem associated with 
the conformal (scalar) mode for which the propagator falls off like $\frac{1}{p^2}$. 
The origin of the scalar mode lies in the Einstein-Hilbert action since gravitational 
Chern-Simons term is in itself conformal invariant so that this topological
term does not affect the behavior of the scalar mode. Thus, the crucial point for proving
the renormalizability of TMG is to find a suitable regularization method for the scalar mode. 

In a recent article \cite{Oda2}, we have solved this problem in the new massive
gravity in three dimensions \cite{Bergshoeff1, Bergshoeff2}. The key observation in the solution
is to add the BRST-invariant Pauli-Fierz mass term in a theory and use it
as an infrared regulator of the scalar mode, and finally take the massless limit
after renormalization procedure. In this article, we will apply this regularization method
to TMG and see that this method is also effective for the proof of renormalizability
of TMG. For that, in the next section, we shall explain the new regularization method
in detail.

\section{BRST-invariant mass term}

Now let us consider the BRST transformations for diffeomorphisms.
The BRST transformations of the metric tensor, ghost, antighost and
Nakanishi-Lautrup auxiliary field are respectively given by
\begin{eqnarray}
\delta_B g_{\mu\nu} &=& - \kappa^3 ( \nabla_\mu c_\nu + \nabla_\nu c_\mu ), 
\nonumber\\
\delta_B c^\mu &=& - \kappa^3 c^\nu \partial_\nu c^\mu, 
\nonumber\\
\delta_B \bar c_\mu &=& i b_\mu,
\nonumber\\
\delta_B b_\mu &=& 0,
\label{BRST}
\end{eqnarray}
where the covariant derivative is defined as usual by
$\nabla_\mu c_\nu = \partial_\mu c_\nu - \Gamma^\lambda_{\mu\nu} c_\lambda$
with the affine connection $\Gamma^\lambda_{\mu\nu}$. It is straightforward to 
prove that the BRST transformations (\ref{BRST}) are off-shell nilpotent.

Next, we expand the metric around a flat Minkowski background $\eta_{\mu\nu}
= diag (-1, 1, 1)$ as usual
\begin{eqnarray}
g_{\mu\nu} = \eta_{\mu\nu} + \kappa h_{\mu\nu}.
\label{G-field}
\end{eqnarray}
With this definition, the gravitational field $h_{\mu\nu}$ has canonical dimension 
of $(mass)^{\frac{1}{2}}$. As a result, the Einstein-Hilbert action and the 
graviton-matter interaction terms have canonical dimensions greater than three, 
which is the origin of unrenormalizability of Einstein's general relativity without 
the higher derivative terms. Later, we will see that the gravitational field 
$h_{\mu\nu}$ has ultraviolet dimension of $(mass)^0$, whose fact leads to 
renormalizability of topologically massive gravity under consideration.
The BRST transformation of the metric in (\ref{BRST}) gives rise to that
of the gravitational field $h_{\mu\nu}$ at the linearized level
\begin{eqnarray}
\delta_B h_{\mu\nu} = - \kappa^2 ( \partial_\mu c_\nu + \partial_\nu c_\mu ), 
\label{h-BRST}
\end{eqnarray}
which is nilpotent up to the order ${\cal{O}}(\kappa^5)$.

Then, we wish to construct a BRST-invariant gravitational field $h'_{\mu\nu}$ at the 
linearized level through the Steuckelberg formalism. To do that, we introduce the Steuckelberg 
vector field $A_\mu$ and define $h'_{\mu\nu}$ as
\begin{eqnarray}
h'_{\mu\nu} = h_{\mu\nu} - \frac{1}{m} ( \partial_\mu A_\nu + \partial_\nu A_\mu ), 
\label{h'}
\end{eqnarray}
and assume the BRST transformation of $A_\mu$ to be \footnote{$\delta_B^2 A_\mu = 0$
up to the order ${\cal{O}}(\kappa^5)$.}
\begin{eqnarray}
\delta_B A_\mu = - \kappa^2 m c_\mu. 
\label{A-BRST}
\end{eqnarray}
Then it is obvious that the $h'_{\mu\nu}$ field is BRST-invariant $\delta_B h'_{\mu\nu} = 0$.

Using this BRST-invariant gravitational field $h'_{\mu\nu}$, let us construct a BRST-invariant
Lagrangian for the Pauli-Fierz mass term  \cite{Fierz} by
\begin{eqnarray}
{\cal{L}}_m^{h'} = - \frac{m^2}{4} ( h'_{\mu\nu} h'^{\mu\nu} - h'^2 ), 
\label{h'-mass 1}
\end{eqnarray}
where $m$ is a constant of mass dimension.
Moreover, we have raised indices by the flat Minkowski metric $\eta^{\mu\nu}$
like $h'^{\mu\nu} = \eta^{\mu\rho} \eta^{\nu\sigma} h'_{\rho\sigma}$ and
$h' = \eta^{\mu\nu} h'_{\mu\nu}$.
When expanded in terms of the definition (\ref{h'}), this Lagrangian reads
\begin{eqnarray}
{\cal{L}}_m^{h'} = - \frac{m^2}{4} ( h_{\mu\nu} h^{\mu\nu} - h^2 )
- \frac{1}{4} F_{\mu\nu}^2 - m (\partial^\nu h_{\mu\nu} - \partial_\mu h) A^\mu, 
\label{h'-mass 2}
\end{eqnarray}
where we have defined $F_{\mu\nu} = \partial_\mu A_\nu - \partial_\nu A_\mu$.

At this stage, we encounter a problem, which is the existence of the ghost
$A^0$. In order to kill this negative norm mode, we need to have an extra
gauge symmetry.  To find it, in particular, let us notice that the last term 
in Eq. (\ref{h'-mass 2}) is not invariant under the usual gauge transformation 
$\delta A_\mu = \partial_\mu \lambda$. Therefore, to remedy this term, we appeal to 
the Steuckelberg formalism again.
If we introduce the Steuckelberg scalar field $\varphi$ via
\begin{eqnarray}
A'_\mu = A_\mu - \frac{1}{m} \partial_\mu \varphi, 
\label{A'}
\end{eqnarray}
the BRST transformations corresponding to new gauge symmetry are determined
such that $\delta_B A'_\mu = 0$ by
\begin{eqnarray}
\delta_B A_\mu &=& \kappa^2 \partial_\mu c,  \nonumber\\
\delta_B \varphi &=& \kappa^2 m c,
\label{c-BRST}
\end{eqnarray}
where $c$ is the new scalar ghost.

As a consequence of the definitions (\ref{h'}) and (\ref{A'}), and
the BRST transformations (\ref{h-BRST}), (\ref{A-BRST}) and (\ref{c-BRST}),
we are eventually led to defining a new BRST-invariant gravitational field 
$\bar h_{\mu\nu}$ by
\begin{eqnarray}
\bar h_{\mu\nu} = h_{\mu\nu} - \frac{1}{m} ( \partial_\mu A_\nu + \partial_\nu A_\mu )
+ \frac{2}{m^2} \partial_\mu \partial_\nu \varphi, 
\label{bar-h}
\end{eqnarray}
and the full BRST transformations at the linearized level by
\begin{eqnarray}
\delta_B h_{\mu\nu} &=& - \kappa^2 ( \partial_\mu c_\nu + \partial_\nu c_\mu ), 
\nonumber\\
\delta_B A_\mu &=& \kappa^2 ( - m c_\mu + \partial_\mu c ),
\nonumber\\
\delta_B \varphi &=& \kappa^2 m c,
\nonumber\\
\delta_B \bar c &=& i b, \ \delta_B b = \delta_B c = 0,
\label{Linearized BRST}
\end{eqnarray}
where the BRST transformations of antighost, Nakanishi-Lautrup field and
ghost are also added. Hence, the new BRST-invariant mass term is 
made out of the BRST-invariant $\bar h_{\mu\nu}$ field as
\begin{eqnarray}
{\cal{L}}_m^{\bar h} &=& - \frac{m^2}{4} ( \bar h_{\mu\nu} \bar h^{\mu\nu} - \bar h^2 )
\nonumber\\
&=& - \frac{m^2}{4} ( h_{\mu\nu} h^{\mu\nu} - h^2 ) - \frac{1}{4} F_{\mu\nu}^2
- ( m A^\mu - \partial^\mu \varphi )(\partial^\nu h_{\mu\nu} - \partial_\mu h). 
\label{bar-h-mass}
\end{eqnarray}

Next, we move on to fixing gauge symmetries. For diffeomorphisms and the
new scalar gauge symmetry respectively, we set up the de Donder gauge and
extended Lorentz gauge conditions
\begin{eqnarray}
\partial_\mu \tilde g^{\mu\nu} \equiv \partial_\mu (\sqrt{-g} g^{\mu\nu}) &=& 0,
\nonumber\\
\Box (\partial_\mu A^\mu - \frac{m}{2} h) &=& 0.
\label{GF}
\end{eqnarray}
The reason why we have selected a higher derivative gauge condition for the
new scalar gauge symmetry will be clarified when we discuss the graviton
propagator in the next section.

The Lagrangian corresponding to the gauge fixing plus FP ghost terms is given 
by the following BRST-exact term:
\begin{eqnarray}
{\cal{L}}_{GF+FP} &=& i \delta_B [ \frac{1}{\kappa^3} \bar c_\nu 
( \partial_\mu \tilde g^{\mu\nu} - \frac{\alpha}{2 \kappa} \eta^{\mu\nu} b_\mu )
+ \kappa^2 \bar c \Box ( \partial_\mu A^\mu - \frac{m}{2} h
- \frac{\beta}{2 \kappa^2} b ) ]
\nonumber\\
&=& \frac{1}{\kappa^3} \tilde g^{\mu\nu} \partial_\mu b_\nu 
+ \frac{\alpha}{2 \kappa^4} \eta^{\mu\nu} b_\mu b_\nu
+ i \partial_\mu \bar c_\nu D_\rho^{\mu\nu} c^\rho 
- \kappa^2 b \Box ( \partial_\mu A^\mu - \frac{m}{2} h )
\nonumber\\
&+& \frac{\beta}{2} b \Box b - i \kappa^4 \bar c \Box^2 c
\nonumber\\
&=& - \frac{1}{2 \alpha \kappa^2} ( \partial_\mu \tilde g^{\mu\nu} )^2
- \frac{1}{2 \beta} \kappa^4 ( \partial_\mu A^\mu - \frac{m}{2} h )
\Box ( \partial_\nu A^\nu - \frac{m}{2} h ) 
\nonumber\\
&+& i \partial_\mu \bar c_\nu D_\rho^{\mu\nu} c^\rho - i \kappa^4 \bar c \Box^2 c, 
\label{GF+FP}
\end{eqnarray}
where $\alpha, \beta$ are gauge parameters and $D_\rho^{\mu\nu} \equiv 
\tilde g^{\mu\sigma} \delta_\rho^\nu \partial_\sigma
+ \tilde g^{\nu\sigma} \delta_\rho^\mu \partial_\sigma 
- \tilde g^{\mu\nu} \partial_\rho - (\partial_\rho \tilde g^{\mu\nu})$.
In the last equality, we have performed path integration over the
Nakanishi-Lautrup fields $b_\mu$ and $b$. 

In this way, we arrive at the gauge fixed, BRST-invariant action
\begin{eqnarray}
S &\equiv& \int d^3 x {\cal{L}}
\nonumber\\
&\equiv& \int d^3 x ( {\cal{L}}_c + {\cal{L}}_m^{\bar h} + {\cal{L}}_{GF+FP} )
\nonumber\\
&=& \int d^3 x [ - \frac{1}{\kappa^2} \sqrt{- g} R  
+ \frac{1}{2 \kappa^2 \mu} \varepsilon^{\lambda\mu\nu} \Gamma_{\lambda\sigma}^\rho
( \partial_\mu \Gamma_{\rho\nu}^\sigma + \frac{2}{3} \Gamma_{\mu\tau}^\sigma
\Gamma_{\nu\rho}^\tau )
\nonumber\\
&-& \frac{m^2}{4} ( h_{\mu\nu} h^{\mu\nu} - h^2 ) - \frac{1}{4} F_{\mu\nu}^2
- ( m A^\mu - \partial^\mu \varphi )(\partial^\nu h_{\mu\nu} - \partial_\mu h)
\nonumber\\
&-& \frac{1}{2 \alpha \kappa^2} ( \partial_\mu \tilde g^{\mu\nu} )^2
- \frac{1}{2 \beta} \kappa^4 ( \partial_\mu A^\mu - \frac{m}{2} h )
\Box ( \partial_\nu A^\nu - \frac{m}{2} h ) 
\nonumber\\
&+& i \partial_\mu \bar c_\nu D_\rho^{\mu\nu} c^\rho - i \kappa^4 \bar c \Box^2 c ].
\label{Final Action}
\end{eqnarray}

To close this section, we should make a comment on the massless limit $m \rightarrow 0$
of the action (\ref{Final Action}). Note that the action (\ref{Final Action})
has a well-defined massless limit and reduces to the form
\begin{eqnarray}
S_{m=0} &=& \int d^3 x [ - \frac{1}{\kappa^2} \sqrt{- g} R  
+ \frac{1}{2 \kappa^2 \mu} \varepsilon^{\lambda\mu\nu} \Gamma_{\lambda\sigma}^\rho
( \partial_\mu \Gamma_{\rho\nu}^\sigma + \frac{2}{3} \Gamma_{\mu\tau}^\sigma
\Gamma_{\nu\rho}^\tau ) - \frac{1}{2 \alpha \kappa^2} ( \partial_\mu \tilde g^{\mu\nu} )^2
+ i \partial_\mu \bar c_\nu D_\rho^{\mu\nu} c^\rho
\nonumber\\
&+& \partial^\mu \varphi(\partial^\nu h_{\mu\nu} - \partial_\mu h) - \frac{1}{4} F_{\mu\nu}^2
- \frac{1}{2 \beta} \kappa^4 \partial_\mu A^\mu \Box \partial_\nu A^\nu 
- i \kappa^4 \bar c \Box^2 c ].
\label{Final Action m = 0}
\end{eqnarray}
This reduced action (\ref{Final Action m = 0}) consists of two parts. The first part
is nothing but the gauge fixed, BRST-invariant action of the classical
action (\ref{TMG}) where the gauge condition for diffeomorphisms 
is the de Donder gauge. 

The second one is a free action made out of only quadratic terms, so that we can simply 
integrate over this part. Indeed, such the terms which do not contain interaction terms are
not relevant at least to the argument of renormalizability.  However, in particular, 
if someone worries the presence of the mixing term between $\varphi$ and $h_{\mu\nu}$ 
in (\ref{Final Action m = 0}),  
it is easy to show that this term can be absorbed into the redefinition of the Nakanishi-Lautrup 
auxiliary field $b_\mu$ at the linearized level if we take slightly modified gauge conditions 
$\partial_\mu (g g^{\mu\nu}) = 0$, which also turns out to lead to the desired property of 
the graviton propagator so that this gauge choice is also admissible. 
Actually, at the linearized order, one has
\begin{eqnarray}
\partial_\mu (g g^{\mu\nu}) = \kappa (\partial_\mu h^{\mu\nu} 
- \partial^\nu h) + \cdots, 
\label{Linear gauge}
\end{eqnarray}
where dots imply the higher-order terms in $\kappa$. 
With the Landau gauge $\alpha = 0$, the gauge fixing and FP ghost Lagrangian for only 
the gauge condition $\partial_\mu (g g^{\mu\nu}) = 0$ takes the form at the
lowest level 
\begin{eqnarray}
{\cal{L}}'_{GF+FP} &=& i \delta_B [\frac{1}{\kappa^3} \bar c_\nu \partial_\mu (g g^{\mu\nu})]
\nonumber\\
&=& - \frac{1}{\kappa^2} b^\mu (\partial^\nu h_{\mu\nu} 
- \partial_\mu h) + i \bar c_\mu (\Box c^\mu - \partial^\mu \partial_\nu c^\nu).
\label{Lagrangian Linear gauge}
\end{eqnarray}
Thus, at the massless limit, up to irrelevant terms for the present argument 
the Lagrangian reads
\begin{eqnarray}
{\cal{L}}'_{m=0} &=& - \frac{1}{\kappa^2} (b^\mu - \kappa^2 \partial^\mu \varphi)
(\partial^\nu h_{\mu\nu} - \partial_\mu h) + i \bar c_\mu (\Box c^\mu 
- \partial^\mu \partial_\nu c^\nu)
\nonumber\\
&\rightarrow& - \frac{1}{\kappa^2} b^\mu (\partial^\nu h_{\mu\nu} 
- \partial_\mu h) + i \bar c_\mu (\Box c^\mu - \partial^\mu \partial_\nu c^\nu),
\label{Lagrangian Linear gauge2}
\end{eqnarray}
where we have redefined $b^\mu - \kappa^2 \partial^\mu \varphi \rightarrow b^\mu$
at the linearized level. In this way, we can nullify the mixing term between 
$\varphi$ and $h_{\mu\nu}$. Let us recall that the theory at hand is independent
of the choice of the gauge condition as well as the gauge parameter, so the 
mixing term can be ignored safely in the present argument.

Finally, of course, this massless limit must be taken after 
the whole renormalization procedure is completed. In this sense, the physical content 
in the present formalism is the same as that of TMG in the massless limit 
although the BRST transformation of the mass term is nilpotent only approximately.

\section{Graviton propagator}

On the basis of the gauge fixed, BRST-invariant action (\ref{Final Action}),
we wish to derive the propagator of the gravitational field $h_{\mu\nu}$
and check that the propagator really has desired features \footnote
{It is true that there are also off-diagonal propagators between gravitons and 
the other fields. However, since there are no interaction terms, it is easy 
to see that they do not make any contribution to the effective action, 
so we can safely neglect them in the argument of renormalizability. Thus
we here consider only the graviton propagator.}.
To this end, it is useful to take account of the spin projection operators 
in $3$ space-time dimensions \cite{Stelle, Nakasone1}. A set of the spin operators 
$P^{(2)}, P^{(1)}, P^{(0, s)}, P^{(0, w)}, P^{(0, sw)}$
and  $P^{(0, ws)}$ form a complete set in the space of second rank symmetric 
tensors and are defined as
\begin{eqnarray}
P^{(2)}_{\mu\nu, \rho\sigma} &=& \frac{1}{2} ( \theta_{\mu\rho} \theta_{\nu\sigma}
+ \theta_{\mu\sigma} \theta_{\nu\rho} ) - \frac{1}{2} \theta_{\mu\nu} \theta_{\rho\sigma},
\nonumber\\  
P^{(1)}_{\mu\nu, \rho\sigma} &=& \frac{1}{2} ( \theta_{\mu\rho} \omega_{\nu\sigma}
+ \theta_{\mu\sigma} \omega_{\nu\rho} + \theta_{\nu\rho} \omega_{\mu\sigma}
+ \theta_{\nu\sigma} \omega_{\mu\rho} ),
\nonumber\\  
P^{(0, s)}_{\mu\nu, \rho\sigma} &=& \frac{1}{2} \theta_{\mu\nu} \theta_{\rho\sigma},
\nonumber\\  
P^{(0, w)}_{\mu\nu, \rho\sigma} &=& \omega_{\mu\nu} \omega_{\rho\sigma},
\nonumber\\  
P^{(0, sw)}_{\mu\nu, \rho\sigma} &=& \frac{1}{\sqrt{2}} \theta_{\mu\nu} \omega_{\rho\sigma},
\nonumber\\  
P^{(0, ws)}_{\mu\nu, \rho\sigma} &=& \frac{1}{\sqrt{2}} \omega_{\mu\nu} \theta_{\rho\sigma}.
\label{Spin projectors}
\end{eqnarray}
Here the transverse operator $\theta_{\mu\nu}$ and the longitudinal operator
$\omega_{\mu\nu}$ are defined as
\begin{eqnarray}
\theta_{\mu\nu} &=& \eta_{\mu\nu} - \frac{1}{\Box} \partial_\mu \partial_\nu 
= \eta_{\mu\nu} - \omega_{\mu\nu}, \nonumber\\  
\omega_{\mu\nu} &=& \frac{1}{\Box} \partial_\mu \partial_\nu.
\label{theta}
\end{eqnarray}
It is straightforward to show that the spin projection operators satisfy 
the orthogonality relations
\begin{eqnarray}
P_{\mu\nu, \rho\sigma}^{(i, a)} P_{\rho\sigma, \lambda\tau}^{(j, b)}
&=& \delta^{ij} \delta^{ab} P_{\mu\nu, \lambda\tau}^{(i, a)},
\nonumber\\  
P_{\mu\nu, \rho\sigma}^{(i, ab)} P_{\rho\sigma, \lambda\tau}^{(j, cd)}
&=& \delta^{ij} \delta^{bc} P_{\mu\nu, \lambda\tau}^{(i, a)},
\nonumber\\  
P_{\mu\nu, \rho\sigma}^{(i, a)} P_{\rho\sigma, \lambda\tau}^{(j, bc)}
&=& \delta^{ij} \delta^{ab} P_{\mu\nu, \lambda\tau}^{(i, ac)},
\nonumber\\  
P_{\mu\nu, \rho\sigma}^{(i, ab)} P_{\rho\sigma, \lambda\tau}^{(j, c)}
&=& \delta^{ij} \delta^{bc} P_{\mu\nu, \lambda\tau}^{(i, ac)},
\label{Orthogonality}
\end{eqnarray}
with $i, j = 0, 1, 2$ and $a, b, c, d = s, w$ and the tensorial relation 
\begin{eqnarray}
[ P^{(2)} + P^{(1)} + P^{(0, s)} + P^{(0, w)} ]_{\mu\nu, \rho\sigma} = 
\frac{1}{2} ( \eta_{\mu\rho}\eta_{\nu\sigma} + \eta_{\mu\sigma}\eta_{\nu\rho} ).
\label{T relation}
\end{eqnarray}

In order to accommodate with gravitational Chern-Simons term,
one has to add two operators to the whole spin projection
operators \cite{Pinheiro, Nakasone2}
\begin{eqnarray}
S_{1 \mu\nu, \rho\sigma} &=& \frac{1}{4} 
\Box ( \varepsilon_{\mu\rho\lambda} \partial_\sigma \omega^\lambda_\nu
+ \varepsilon_{\mu\sigma\lambda} \partial_\rho \omega^\lambda_\nu
+ \varepsilon_{\nu\rho\lambda} \partial_\sigma \omega^\lambda_\mu
+ \varepsilon_{\nu\sigma\lambda} \partial_\rho \omega^\lambda_\mu ),
\nonumber\\  
S_{2 \mu\nu, \rho\sigma} &=& - \frac{1}{4} 
\Box ( \varepsilon_{\mu\rho\lambda} \eta_{\sigma\nu}
+ \varepsilon_{\mu\sigma\lambda} \eta_{\rho\nu}
+ \varepsilon_{\nu\rho\lambda} \eta_{\sigma\mu}
+ \varepsilon_{\nu\sigma\lambda} \eta_{\rho\mu} ) \partial^\lambda.
\label{Spin projectors2}
\end{eqnarray}
These operators together with the spin projection operators satisfy the
following relations
\begin{eqnarray}
S_1 S_1 &=& \frac{1}{4} \Box^3 P^{(1)}, \nonumber\\
S_1 S_2 &=& S_2 S_1 = - \frac{1}{4} \Box^3 P^{(1)}, \nonumber\\
S_2 S_2 &=& \Box^3 ( P^{(2)} + \frac{1}{4} P^{(1)} ), \nonumber\\
P^{(1)} S_1 &=& S_1 P^{(1)} = S_1, \nonumber\\
P^{(1)} S_2 &=& S_2 P^{(1)} = - S_1, \nonumber\\
P^{(2)} S_2 &=& S_2 P^{(2)} = S_1 + S_2,
\label{Spin projectors3}
\end{eqnarray}
where the matrix indices on operators are to be understood.

It is then straightforward to extract the quadratic fluctuations in 
$h_{\mu\nu}$ from each term in the action (\ref{Final Action}) and express them
in terms of the spin projection operators and $S$ operators:
\begin{eqnarray}
{\cal{L}}_{EH} &\equiv& - \frac{1}{\kappa^2} \sqrt{- g} R 
\nonumber\\
&=& - \frac{1}{4} h^{\mu\nu} [ P^{(2)} - P^{(0, s)} ]_{\mu\nu, \rho\sigma}
\Box h^{\rho\sigma},
\nonumber\\
{\cal{L}}_{GCS} &\equiv& \frac{1}{2 \kappa^2 \mu} \varepsilon^{\lambda\mu\nu} 
\Gamma_{\lambda\sigma}^\rho ( \partial_\mu \Gamma_{\rho\nu}^\sigma + \frac{2}{3} 
\Gamma_{\mu\tau}^\sigma \Gamma_{\nu\rho}^\tau )
\nonumber\\
&=& \frac{1}{4 \mu} h^{\mu\nu} ( S_1 + S_2 )_{\mu\nu, \rho\sigma} h^{\rho\sigma},
\nonumber\\
{\cal{L}}_{PF} &\equiv& - \frac{m^2}{4} ( h_{\mu\nu} h^{\mu\nu} - h^2 )
\nonumber\\
&=& - \frac{m^2}{4} h^{\mu\nu} [ P^{(2)} + P^{(1)} - P^{(0, s)} 
- \sqrt{2} ( P^{(0, sw)} + P^{(0, ws)} ) ]_{\mu\nu, \rho\sigma} h^{\rho\sigma},
\nonumber\\
{\cal{L}}_\alpha &\equiv& - \frac{1}{2 \alpha \kappa^2} ( \partial_\mu \tilde g^{\mu\nu} )^2
\nonumber\\
&=& \frac{1}{2 \alpha} h^{\mu\nu} [ \frac{1}{2} P^{(1)} + \frac{1}{2} P^{(0, s)} 
+ \frac{1}{4} P^{(0, w)} - \frac{1}{2 \sqrt{2}} 
( P^{(0, sw)} + P^{(0, ws)} )]_{\mu\nu, \rho\sigma} \Box h^{\rho\sigma},
\nonumber\\
{\cal{L}}_\beta &\equiv& - \frac{1}{2 \beta} \kappa^4 \frac{m^2}{4} h \Box h
\nonumber\\
&=& - \frac{1}{2 \beta} \kappa^4 \frac{m^2}{4} h^{\mu\nu} [ 2 P^{(0, s)} + P^{(0, w)} 
+ \sqrt{2} ( P^{(0, sw)} + P^{(0, ws)} )]_{\mu\nu, \rho\sigma} \Box h^{\rho\sigma}.
\label{Lagrangian}
\end{eqnarray}

Using the relations (\ref{Lagrangian}), 
the quadratic part in $h_{\mu\nu}$ of the action (\ref{Final Action}) is expressed
in terms of the spin projection operators and $S$ operators
\begin{eqnarray}
S = \int d^3 x \frac{1}{2} h^{\mu\nu} {\cal{P}}_{\mu\nu, \rho\sigma} 
h^{\rho\sigma},
\label{Quadratic Action}
\end{eqnarray}
where ${\cal{P}}_{\mu\nu, \rho\sigma}$ is defined as
\begin{eqnarray}
{\cal{P}}_{\mu\nu, \rho\sigma} &=& [ - \frac{1}{2} ( \Box + m^2 ) P^{(2)} 
+ \frac{1}{2} ( \frac{1}{\alpha} \Box - m^2 ) P^{(1)} 
+ \frac{1}{2} \{ ( 1 + \frac{1}{\alpha} - \frac{1}{\beta} \kappa^4 m^2 ) \Box + m^2 \} P^{(0, s)}
\nonumber\\
&+& \frac{1}{4} ( \frac{1}{\alpha} - \frac{1}{\beta} \kappa^4 m^2 ) \Box P^{(0, w)}
- \frac{1}{2 \sqrt{2}} \{ ( \frac{1}{\alpha} + \frac{1}{\beta} \kappa^4 m^2 ) \Box - 2 m^2 \}
( P^{(0, sw)} + P^{(0, ws)} ) 
\nonumber\\
&+& \frac{1}{2 \mu} ( S_1 + S_2 ) ]_{\mu\nu, \rho\sigma}.
\label{P}
\end{eqnarray}
Then, the propagator of $h_{\mu\nu}$ is defined in a standard way by
\begin{eqnarray}
<0| T (h_{\mu\nu}(x) h_{\rho\sigma}(y)) |0>
= i {\cal{P}}_{\mu\nu, \rho\sigma}^{-1} \delta^{(3)}(x-y).
\label{Propagator}
\end{eqnarray}
With the help of the relations (\ref{Orthogonality}), (\ref{T relation})
and (\ref{Spin projectors3}), 
the inverse of the operator $\cal{P}$ is easily calculated to be
\begin{eqnarray}
{\cal{P}}_{\mu\nu, \rho\sigma}^{-1} 
&=& [ x_1 P^{(2)} + x_2 P^{(1)} + x_3 P^{(0, s)} + x_4 P^{(0, w)}
+ x_5 ( P^{(0, sw)} + P^{(0, ws)} ) 
\nonumber\\
&+& x_6 (S_1 + S_2) ]_{\mu\nu, \rho\sigma},
\label{P-inv}
\end{eqnarray}
where $x_i (i = 1, 2, \cdots, 6)$ are given by
\begin{eqnarray}
x_1 &=& \frac{2 \mu ( \Box + m^2 )}{\frac{1}{\mu} \Box^3 - \mu (\Box + m^2)^2},
\nonumber\\
x_2 &=& \frac{2}{\frac{1}{\alpha} \Box - m^2},
\nonumber\\
x_3 &=& \frac{ 2 ( \frac{1}{\alpha} - \frac{1}{\beta} \kappa^4 m^2 ) \Box}{I},
\nonumber\\
x_4 &=& \frac{ 4 [ ( 1 + \frac{1}{\alpha} - \frac{1}{\beta} \kappa^4 m^2 ) \Box
+ m^2 ] }{I},
\nonumber\\
x_5 &=& \frac{ 2 \sqrt{2} [ ( \frac{1}{\alpha} + \frac{1}{\beta} \kappa^4 m^2 ) \Box
- 2 m^2 ] }{I},
\nonumber\\
x_6 &=& \frac{2}{\frac{1}{\mu} \Box^3 - \mu (\Box + m^2)^2},
\label{x}
\end{eqnarray}
with $I$ being defined by
\begin{eqnarray}
I = [ \frac{1}{\alpha} - \frac{1}{\beta} ( 1 + \frac{4}{\alpha} ) \kappa^4 m^2 ] \Box^2
+ ( \frac{5}{\alpha} + \frac{3}{\beta} \kappa^4 m^2 ) m^2 \Box - 4 m^4.
\label{I}
\end{eqnarray}

These expressions provide us how to choose the gauge parameters $\alpha, \beta$ 
in order to have the desired graviton propagator. Before doing so, let us
recall where the difficult point is in the proof of renormalizability of TMG \cite{Deser3}.
Since gravitational Chern-Simons term includes three derivaives, we expect that 
the graviton propagator would fall off like $\frac{1}{p^3}$ for large momenta. In fact, 
this $\frac{1}{p^3}$ behavior turns out to be required for power counting renormalizability of TMG. 
However, in particular, the propagator of the scalar excitation ($x_3$) behaves as 
$\frac{1}{p^2}$ and thus decreases more slowly in the large momentum limit.
The reason why the scalar mode behaves like so is simple. This part of the
propagator comes from the conformal mode in the Einstein-Hilbert action and
is not affected by the higher derivative gravitational Chern-Simons term
since the gravitational Chern-Simons term is conformally invariant. 

This thorny problem is resolved in the present formalism as follows:
First, let us notice that the coefficients $x_2, x_3, x_4$ and $x_5$
depend on the gauge parameters $\alpha, \beta$ whereas $x_1$ and $x_6$
are independent of them. Second, the troublesome spin 0 component $x_3$
projected by $P^{(0, s)}$ can be vanished by selecting the condition
$\beta = \alpha \kappa^4 m^2$. Third, the spin 1 component $x_2$
projected by $P^{(1)}$ vanishes when we take the limit $\alpha \rightarrow 0$.
In other words, we choose the gauge parameters to be
\begin{eqnarray}
\beta = \alpha \kappa^4 m^2 \rightarrow 0.
\label{Gauge parameters}
\end{eqnarray}
Finally, it is easy to see that with this condition (\ref{Gauge parameters})
on the gauge parameters, the other gauge-variant terms $x_4$ and $x_5$
become zero as well.

Consequently, with the condition (\ref{Gauge parameters}) we have the propagator of 
the graviton whose essential part is controlled by 
\begin{eqnarray}
{\cal{P}}_{\mu\nu, \rho\sigma}^{-1} 
= [ \frac{2 \mu ( \Box + m^2 )}{\frac{1}{\mu} \Box^3 - \mu (\Box + m^2)^2} P^{(2)}
+ \frac{2}{\frac{1}{\mu} \Box^3 - \mu (\Box + m^2)^2} (S_1 + S_2) ]_{\mu\nu, \rho\sigma},
\label{P-inv2}
\end{eqnarray}
An interesting feature is that this propagator is described entirely by 
the transverse operator $\theta_{\mu\nu}$ so it vanishes when multiplied by the momenta $p^\mu$. 

We will soon realize that this propagator damps cubically like $\frac{1}{p^3}$ for large momenta,
instead of the quadratic one $\frac{1}{p^2}$ as was obtained by the Einstein-Hilbert action,
since the S operators involve the momentum factor $p^2 p^\mu$. Thus, the dimension of the
gravitational field $h_{\mu\nu}$, concerning its ultraviolet behavior, should be assigned
to be $(mass)^0$ in place of the original canonical one $(mass)^{\frac{1}{2}}$.
With this assignment of the dimension, interaction terms existing in the Einstein-Hilbert
action have the dimension $(mass)^2$ solely coming from two derivatives and the
index $\delta_{EH} = 2 - 3 = - 1$, thereby implying that they are super-renormalizable.
Similarly, interaction terms in gravitational Chern-Simons term have the dimension
$(mass)^3$ and hence the index $\delta_{GCS} = 3 - 3 = 0$, so they are of marginally
renormalizable type.

As a final remark, it is worthwhile to ask ourselves what has become of one dynamical degree
of freedom associated with the scalar mode since we know that the number of dynamical degrees
of freedom remains unchanged in perturbation theory. The answer lies in the fact 
that there appears the propagator for the Steuckelberg field $A_\mu$, which has one dynamical 
degree of freedom owing to gauge invariance. In other words, because of the BRST-invariant regulator 
the massless pole $\frac{1}{p^2}$ of the scalar mode is changed to that of the Steuckelberg 
field $A_\mu$.

\section{Superficial degree of divergence and ultraviolet regulator}

We now turn our attention to the analysis of structure of the divergences.
To this aim, let us first add sources $K_{\mu\nu}$ (anti-commuting, ghost
number = $-1$, dimension = $\frac{3}{2}$), $L_\mu$ (commuting, ghost number = $-2$, 
dimension = $1$), $M_\mu$ (anti-commuting, ghost number = $-1$, dimension = $1$)
and $N$ (anti-commuting, ghost number = $-1$, dimension = $1$) for the BRST 
transformations of $\tilde g^{\mu\nu}$, $c^\mu$, $A^\mu$ and $\varphi$ to the 
action (\ref{Final Action}), respectively \footnote{Here for simplicity we have
regarded $\tilde g^{\mu\nu}$ as a basic gravitational field.}:
\begin{eqnarray}
\tilde S &\equiv& \int d^3 x \tilde \Sigma 
\nonumber\\
&=& \int d^3 x [ {\cal{L}}
+ \frac{1}{\kappa^3} K_{\mu\nu} \delta_B \tilde g^{\mu\nu} 
+ \frac{1}{\kappa^3} L_\mu \delta_B c^\mu 
+ \frac{1}{\kappa^3} M_\mu \delta_B A^\mu 
+ \frac{1}{\kappa^3} N \delta_B \varphi ]
\nonumber\\
&=& \int d^3 x [ - \frac{1}{\kappa^2} \sqrt{- g} R  
+ \frac{1}{2 \kappa^2 \mu} \varepsilon^{\lambda\mu\nu} \Gamma_{\lambda\sigma}^\rho
( \partial_\mu \Gamma_{\rho\nu}^\sigma + \frac{2}{3} \Gamma_{\mu\tau}^\sigma
\Gamma_{\nu\rho}^\tau )
\nonumber\\
&-& \frac{m^2}{4} ( h_{\mu\nu} h^{\mu\nu} - h^2 ) - \frac{1}{4} F_{\mu\nu}^2
- ( m A^\mu - \partial^\mu \varphi )(\partial^\nu h_{\mu\nu} - \partial_\mu h)
\nonumber\\
&-& \frac{1}{2 \alpha \kappa^2} ( \partial_\mu \tilde g^{\mu\nu} )^2
- \frac{1}{2 \beta} \kappa^4 ( \partial_\mu A^\mu - \frac{m}{2} h )
\Box ( \partial_\nu A^\nu - \frac{m}{2} h ) 
+ ( K_{\mu\nu} + i \partial_\mu \bar c_\nu ) D_\rho^{\mu\nu} c^\rho 
\nonumber\\
&-& i \kappa^4 \bar c \Box^2 c - L_\mu c^\nu \partial_\nu c^\mu 
+ \frac{1}{\kappa} M_\mu ( -m c^\mu + \partial^\mu c ) + \frac{m}{\kappa} N c ].
\label{BRST-Action with sources}
\end{eqnarray}

Next, based on this action, let us consider the superficial degree of divergence
for 1PI (one particle irreducible) Feynman diagrams. 
Then, it is convenient to introduce the following notation:
$n_R =$ number of graviton vertices with two derivatives, $n_G =$ number of ghost vertices,
$n_K =$ number of $K$-graviton-ghost vertices, $n_L =$ number of $L$-ghost-ghost vertices,
$I_G =$ number of internal ghost propagators and $I_E =$ number of internal
graviton propagators. Note that the fields $A_\mu, \varphi, c$ and $\bar c$ are 
free so we can exclude such the fields from the counting of the superficial degree of 
divergence. Using this fact and the above notation, the superficial degree of divergence 
for an arbitrary Feynman diagram $\gamma$ can be easily calculated to be
\begin{eqnarray}
\omega(\gamma) &\equiv& \sum n_i d_i + (3-3) I_E + (3-2) I_G
- 3 (\sum n_i - 1)
\nonumber\\
&=& 3 - n_R - n_G - 2 n_K - 2 n_L + I_G,
\label{SDD1}
\end{eqnarray}
where $d_i$ denotes the number of derivatives in the interaction terms.
Here we have made use of the fact that the graviton propagator behaves like $p^{-3}$
for large momenta as mentioned in the previous section.

Furthermore, using the relation
\begin{eqnarray}
2 n_G + 2 n_L + n_K = 2 I_G + E_c + E_{\bar c},
\label{Ghost relation}
\end{eqnarray}
with the notation that $E_c =$ number of external ghosts $c^\mu$ and
$E_{\bar c} =$ number of external antighosts $\bar c_\mu$,
Eq. (\ref{SDD1}) is cast to the form
\begin{eqnarray}
\omega(\gamma) = 3 - n_R - \frac{3}{2} n_K -  n_L 
- \frac{1}{2} E_c -  \frac{1}{2} E_{\bar c}.
\label{SDD2}
\end{eqnarray}
Accordingly, we reach the conclusion
\begin{eqnarray}
\omega(\gamma) \le 3,
\label{SDD3}
\end{eqnarray}
which indicates that the bound on the superficial degree of divergence
for 1PI diagrams $\gamma$ is cubic to all orders, and the theory is power-counting
renormalizable. 

Next, we have to take into consideration the ultraviolet regularization
in order to deal with divergent integrals. Here we meet another difficulty. 
Because TMG includes the Levi-Civita tensor density explicitly, we
cannot use dimensional regularization beyond one-loop. Furthermore,
it turned out that covariant, higher derivative regularization spoils
the argument of renormalizability when the unusual parametrization
of the metric tensor is used \cite{Deser3}.

On the other hand, with the usual parametrization of the metric (\ref{G-field}), 
this problem can be solved by using a covariant, higher derivative regularization 
as will be explained shortly. Note that there are two criteria of selecting a suitable 
regulator. One criterion is that the regulator should push all divergences to
only one-loop where we can make use of powerful dimensional regularization. 
Another criterion is that regularization should not generate 
the graviton propagator, in particular, the propagator of the scalar
mode, which falls off like $\frac{1}{p^2}$ for large momenta.
It is of interest that the simplest candidate is provided from the new massive
gravity \cite{Bergshoeff1, Bergshoeff2}. As mentioned explicitly in Refs. \cite{Nakasone2, Oda2},
at the quadratic order in $\kappa^2$, a salient feature of the specific combination of 
$R^2$ terms in the new massive gravity is the disappearance of the spin projection operator 
corresponding to the spin 0 scalar mode:
\begin{eqnarray}
\sqrt{- g} ( R_{\mu\nu} R^{\mu\nu} - \frac{3}{8} R^2 )
\approx \frac{\kappa^2}{4} h^{\mu\nu} P^{(2)}_{\mu\nu, \rho\sigma}
\Box^2 h^{\rho\sigma},
\label{Peculiar relation}
\end{eqnarray}
thereby the higher derivative sector becoming conformally invariant \cite{Deser4}. 
 
Using this observation, we shall propose the following covariant, higher deivative 
regulator satisfying the above criteria
\begin{eqnarray}
S_\Lambda &=& \frac{1}{\kappa^2 \Lambda^6} \int d^3 x \sqrt{- g} 
( R_{\mu\nu} \Box_g^2 R^{\mu\nu} - \frac{3}{8} R \Box_g^2 R )
\nonumber\\
&\approx& \frac{1}{4 \Lambda^6} \int d^3 x h^{\mu\nu} P^{(2)}_{\mu\nu, \rho\sigma}
\Box^4 h^{\rho\sigma},
\label{Regulator}
\end{eqnarray}
where $\Box_g^2 \equiv g^{\mu\nu} g^{\rho\sigma} \nabla_\mu \nabla_\nu \nabla_\rho 
\nabla_\sigma$. Note that $S_\Lambda$ has the graviton propagator damping like
$\frac{1}{p^8}$ for large momenta and vertices with eight derivatives.
Indeed, with the condition (\ref{Gauge parameters}), ${\cal{P}}_{\mu\nu, \rho\sigma}^{-1}$
is now replaced with 
\begin{eqnarray}
{\cal{P}}_{\mu\nu, \rho\sigma}^{-1} 
= [ \frac{2 \mu ( - \frac{1}{\Lambda^6} \Box^4 + \Box + m^2 )}
{\frac{1}{\mu} \Box^3 - \mu ( - \frac{1}{\Lambda^6} \Box^4 + \Box + m^2)^2} P^{(2)}
+ \frac{2}{\frac{1}{\mu} \Box^3 - \mu ( - \frac{1}{\Lambda^6} \Box^4 + \Box + m^2)^2} 
(S_1 + S_2) ]_{\mu\nu, \rho\sigma},
\label{P-inv3}
\end{eqnarray}
which certainly falls off like $\frac{1}{p^8}$ for large momenta.

Repeating the calculation of the superficial degree of divergence with this regulator,
we now find that
\begin{eqnarray}
\omega(\gamma) = 3 - n_R - 5 (I_E - n_\Lambda) - \frac{3}{2} n_K -  n_L 
- \frac{1}{2} E_c -  \frac{1}{2} E_{\bar c},
\label{SDD4}
\end{eqnarray}
where $n_\Lambda$ is number of graviton vertices with eight derivatives.
This $\omega(\gamma)$ also satisfies the inequality (\ref{SDD3}) since $I_E \geq n_\Lambda$
for 1PI diagrams. The topological relation among numbers of loops,
internal lines and vertices yields the equation
\begin{eqnarray}
L = I_E + I_G - n_R - n_{GCS} - n_\Lambda - n_G - n_K -  n_L + 1, 
\label{Top relation}
\end{eqnarray}
with $n_{GCS}$ being number of graviton vertices with three derivatives,
so we can rewrite the superficial degree of divergence in terms of loop
number $L$
\begin{eqnarray}
\omega(\gamma) = 8 - 5 L - 6 n_R -4 n_K -  n_L - 5 n_{GCS}
- 3 E_c -  3 E_{\bar c}.
\label{SDD5}
\end{eqnarray}
Hence, since $\omega(\gamma)$ becomes negative for $L \geq 2$ while it
has a possibility of becoming positive for $L = 1$, we have divergent 
integrals only in the one-loop amplitudes where we can use dimensional 
regularization to evaluate such divergent ones.

Although we do not calculate divergent terms in this article explicitly,
it is sufficient to prove renormalizability of TMG by using the analysis
argued so far. Since the divergent part is local, BRST-invariant (that is,
diffeomorphism-invariant), and of dimension $3$ at most, the possible 
form of the divergent counter-terms reads
\begin{eqnarray}
S_{counter} &=& \int d^3 x [ a_1 \sqrt{- g} + a_2 \sqrt{- g} R  
+ a_3 \varepsilon^{\lambda\mu\nu} \Gamma_{\lambda\sigma}^\rho
( \partial_\mu \Gamma_{\rho\nu}^\sigma + \frac{2}{3} \Gamma_{\mu\tau}^\sigma
\Gamma_{\nu\rho}^\tau ) ],
\label{Counter-term}
\end{eqnarray}
where $a_i (i = 1, 2, 3)$ are divergent coefficients. These divergences can
be absorbed by renormalizations of the Newton's constant and the parameter $\mu$ 
in front of gravitational Chern-Simons term, and by adding the cosmological
counter-term. We have thus completed the proof of renormalizability of
topologically massive gravity in three dimensions.

\section{Discussions}

In this article, we have presented a proof of renormalizability of topologically 
massive gravity (TMG) in three dimensions \cite{Deser1, Deser2}.
This problem has been already studied in Refs. \cite{Deser3, Kleppe},
but their proofs were incomplete, unfortunately.  One of the differences between their approach
and ours is that they have used the unconventional parametrization of the
metric tensor whereas we have done the usual paramerization. 
The two (closely related) problems, those are, the one associated with the scalar 
propagator and the other being the ultraviolet regulator, are resolved
in our approach. In particular, we should notice that our proof relies on the existence 
of both the BRST symmetry and the BRST-invariant infrared regulator.

Furthermore, we have proposed the ultraviolet regulator whose form is motivated
by the new massive gravity in three dimensions \cite{Bergshoeff1, Bergshoeff2}.
Here it is worthwhile to point out that TMG and the new massive gravity share
one common feature, that is, conformal invariance in the higher derivative sector.
We think that this common feature might be an essential ingredient of renormalizability
of quantum gravity in three dimensions.

Since it has been already shown that TMG is an interactive and unitary theory for 
the massive graviton of spin $+2$ or $-2$, our proof of renormalizability
supports that this theory is a satisfactory example of quantum gravity though it is 
formulated only in three dimensions.

\vs 1   

\end{document}